\documentclass[sigconf, nonacm]{acmart}

\usepackage{graphicx} 
\usepackage{enumitem}
\usepackage{tikz}

\newcommand\vldbdoi{XX.XX/XXX.XX}
\newcommand\vldbpages{XXX-XXX}
\newcommand\vldbvolume{14}
\newcommand\vldbissue{1}
\newcommand\vldbyear{2020}
\newcommand\vldbauthors{\authors}
\newcommand\vldbtitle{\shorttitle} 
\newcommand\vldbavailabilityurl{https://github.com/erbench/erbench}
\newcommand\vldbpagestyle{plain} 

\begin{document}
\title{Agentic ER: The next frontier in Entity Resolution [Vision]}

\author{George Papadakis}
\orcid{0000-0002-3878-5988}
\affiliation{%
  \institution{National and Kapodistrian University of Athens}
  \city{Athens}
  \country{Greece}
}
\email{gpapadis@di.uoa.gr}

\author{Panos Korovesis}
\orcid{XXXXXX}
\affiliation{%
  \institution{National and Kapodistrian University of Athens}
  \city{Athens}
  \country{Greece}
}
\email{panosk@di.uoa.gr}

\author{Manolis Koubarakis}
\orcid{0000-0002-1954-8338}
\affiliation{%
  \institution{National and Kapodistrian University of Athens}
  \city{Athens}
  \country{Greece}
}
\email{koubarak@di.uoa.gr}

\author{Themis Palpanas}
\orcid{0000-0002-8031-0265}
\affiliation{%
  \institution{Université Paris Cité}
  \city{Paris}
  \country{France}
  }
\email{themis@mi.parisdescartes.fr}

\begin{abstract}

Entity Resolution (ER) is a fundamental problem in data management, playing a critical role in tasks like data cleaning and knowledge graph construction. The existing ER approaches range from traditional rule-based to deep learning techniques and LLM-based methods, but typically operate under a ``passive paradigm'', as duplicates are detected through static, one-shot similarity computations. Such approaches fail to capture the inherently uncertain and context-dependent nature of real-world ER tasks, especially in data lakes with streaming content in heterogeneous formats such as CSV files, JSON files, RDF dumps, and free text. In such settings, resolving ambiguity often requires iterative evidence gathering, reasoning across multiple sources, even selective human involvement. To cover this gap, we advocate a paradigm shift from passive to Agentic ER, which frames ER as a sequential decision-making process that is performed by autonomous agents. These agents actively plan ER strategies, acquire external evidence, decide when to query additional sources or humans, and optimize trade-offs between accuracy, cost, and latency. We formalize Agentic ER as a decision-theoretic problem, we propose a reference architecture, we identify core research challenges, and outline new evaluation dimensions tailored to agentic behavior. By introducing Agentic ER, we aim to establish a new research direction at the intersection of data management and intelligent agents.
\end{abstract}

\maketitle

\pagestyle{\vldbpagestyle}
\begingroup\small\noindent\raggedright\textbf{PVLDB Reference Format:}\\
\vldbauthors. \vldbtitle. PVLDB, \vldbvolume(\vldbissue): \vldbpages, \vldbyear.\\
\href{https://doi.org/\vldbdoi}{doi:\vldbdoi}
\endgroup
\begingroup
\renewcommand\thefootnote{}\footnote{\noindent
This work is licensed under the Creative Commons BY-NC-ND 4.0 International License. Visit \url{https://creativecommons.org/licenses/by-nc-nd/4.0/} to view a copy of this license. For any use beyond those covered by this license, obtain permission by emailing \href{mailto:info@vldb.org}{info@vldb.org}. Copyright is held by the owner/author(s). Publication rights licensed to the VLDB Endowment. \\
\raggedright Proceedings of the VLDB Endowment, Vol. \vldbvolume, No. \vldbissue\ %
ISSN 2150-8097. \\
\href{https://doi.org/\vldbdoi}{doi:\vldbdoi} \\
}\addtocounter{footnote}{-1}\endgroup

\ifdefempty{\vldbavailabilityurl}{}{
\vspace{.3cm}
}

\section{Introduction}

Entity Resolution (ER) constitutes the core task of identifying records that refer to the same real-world entity across or within data sources \cite{DBLP:books/daglib/0030287}. In fact, ER is a critical enabler for data integration, data cleaning, knowledge graph construction, and downstream analytics \cite{DBLP:journals/pvldb/0001LSDT20}. From consolidating customer records across enterprise databases to interlinking entities in web-scale knowledge graphs, the quality of ER is critical for data-driven systems \cite{DBLP:journals/csur/ChristophidesEP21}. In modern data lakes, this challenge is compounded by heterogeneous data formats (from CSV files and relational databases to JSON files, RDF dumps, and free text), while the entities must be resolved incrementally as new data streams arrive \cite{DBLP:conf/data/BouabdelliAHH25,DBLP:journals/tkde/HaiKQJ23}.

Despite decades of research, most ER approaches adhere to a \textit{passive paradigm}. Most existing methods follow a fixed pipeline consisting of blocking \cite{DBLP:journals/pvldb/PaulsenGD23,DBLP:journals/corr/abs-2304-12329,DBLP:journals/corr/abs-2202-12521}, pairwise matching \cite{DBLP:journals/pvldb/DongR18,Papadakis2021morgan}, and clustering \cite{DBLP:journals/vldb/PapadakisETHC23,DBLP:journals/pvldb/HassanzadehCML09}, where decisions are made based on static similarity functions or learned classifiers. More recent approaches leverage deep learning \cite{DBLP:journals/corr/abs-2207-04122,DBLP:journals/pvldb/Thirumuruganathan21,DBLP:conf/www/ChenSZ20,DBLP:journals/pvldb/0001LSDT20} and large language models (LLMs) \cite{DBLP:conf/www/LiLHZSC24,DBLP:conf/coling/WangCLCHSWZ25} to improve matching accuracy, yet they largely retain the same underlying assumption: ER is performed as a \textit{one-shot decision problem}, where each pair (or group) of records is evaluated independently, without iterative reasoning or adaptive behavior.

However, real-world ER is rarely a one-shot task. For example, suppose that we aim to resolve whether the following two records refer to the same researcher: \texttt{Record A: ``J. Smith, University of Athens, works on data integration''} and \texttt{Record B: ``John Smith, smith@di.uoa.gr, publications in entity matching''}. An existing ER system would compute similarity scores over names, affiliations, and keywords, and based on them, it would output a match or a non-match decision.

Such passive ER approaches are not suitable for modern, real-world settings, which involve complex, dynamic data lakes with continuously evolving, heterogeneous content spanning structured, semi-structured, and unstructured formats, with high levels of noise and incomplete information \cite{DBLP:conf/data/BouabdelliAHH25,DBLP:journals/tkde/HaiKQJ23}. More specifically, existing ER solutions suffer from the following drawbacks:

\begin{enumerate}[leftmargin=*]
    \item \textit{Lack of Sequential Reasoning.}
    They treat ER as a primarily classification problem, ignoring the fact that resolving ambiguous entities often requires \textit{multi-step reasoning}. E.g., intermediate decisions, such as identifying potential ambiguities or inconsistencies, do not trigger additional actions or refinement steps.
    \item \textit{No Adaptive Evidence Acquisition.}
    They assume a \textit{closed-world setting}, where all relevant information is available upfront. However, in practice critical evidence may reside in external sources such as web pages or knowledge graphs.
    Yet, existing ER solutions lack mechanisms to decide \textit{when} additional evidence is needed, to determine \textit{which sources} to query, and to incorporate newly acquired information into the ER process.
    \item \textit{No Interaction or Feedback Integration.}
    They operate in isolation, without the ability to interact with users or integrate evidence from other ER techniques during the resolution process. Active learning techniques allow selective labeling during training, but they do not support \textit{interactive decision-making at inference time}, i.e., they cannot ask clarification questions, resolve uncertainty through targeted queries, or incorporate feedback dynamically.
    \item \textit{Lack of Cost Awareness.}
    They focus primarily on maximizing effectiveness, without explicitly modeling the \textit{cost of computation, data access, or human interaction}. In real-world applications, however, these costs are often critical constraints. For instance, querying external sources may incur latency or monetary cost, human input is expensive and limited, and large-scale comparisons must be carefully budgeted. Without cost-aware decision-making, existing solutions cannot optimize trade-offs between effectiveness and resource usage, further limiting their applicability in large-scale or time-sensitive settings.
\end{enumerate}

\begin{figure}[t]
\centerline{\includegraphics[width=0.5\textwidth]{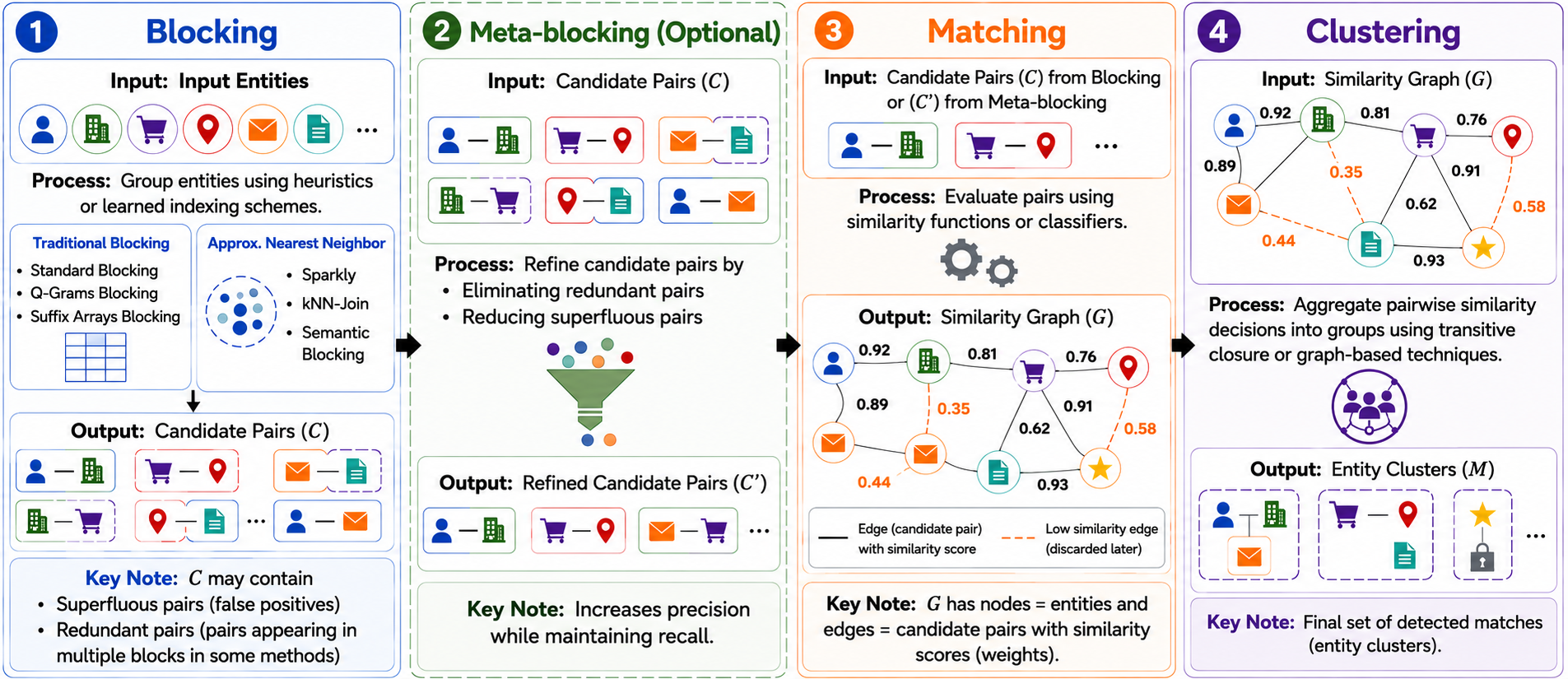}}
\vspace{-8pt}
\caption{The end-to-end pipeline of Passive ER.}
\vspace{-12pt}
\label{fig:pipeline}
\end{figure}

Taken together, these limitations reveal a fundamental gap in current ER methodologies. This observation motivates the need for a new paradigm that enables \textit{multi-step reasoning, adaptive evidence acquisition, interaction, and cost-aware optimization}. 

To cover this gap, we argue for a fundamental shift in how ER is conceptualized and implemented. We propose \textbf{Agentic ER}, a new paradigm in which ER is framed as a sequential decision-making process carried out by an autonomous agent. This paradigm leverages recent advances in LLMs and autonomous agents that can plan, reason, and interact with external tools.

In more detail, an Agentic ER solution does not simply classify pairs of records. Instead, it leverages agents that \textit{plan} a sequence of actions to resolve entities, \textit{acquire evidence} from internal and external sources,  \textit{adapt} their strategy based on intermediate observations, \textit{interact} with humans or other systems when needed, and \textit{optimize} trade-offs between accuracy, cost, and latency. This perspective shifts the focus from \textit{similarity computation} to \textit{goal-directed reasoning}, enabling ER systems to operate effectively in complex, open-world environments.


Overall, this work advocates Agentic ER as a new research direction, which lays the foundation for the next generation of ER solutions that are capable of operating in complex, real-world environments. As contributions, we formalize Agentic ER, we propose a reference architecture, we identify core research challenges, and we outline new evaluation dimensions.

\section{Passive ER solutions}

The core notion in ER is the \textit{entity profile}, i.e., the description of a real-world object. We define it as a set of name-value pairs that are associated with a distinct ID \cite{DBLP:journals/corr/abs-2304-12329}. This simple model is flexible enough to uniformly accommodate the main data formats in a common data lake with streaming content, such as CSV files, relational databases, JSON files, RDF dumps, knowledge graphs, free text etc.

Two are the main forms of Entity Resolution in the literature~\cite{DBLP:journals/csur/ChristophidesEP21}:
\begin{enumerate}[leftmargin=*]
    \item \textit{Clean-Clean ER} or \textit{Record Linkage} receives as input two entity collections, $\mathcal{E}_1$ and $\mathcal{E}_2$, that are individually duplicate-free (i.e., clean), but overlapping. As output, it returns as the set of duplicate entities $\mathcal{D} = \{<e_i, e_j> : e_i \equiv e_j \wedge e_i \in \mathcal{E}_1 \wedge e_j \in \mathcal{E}_2 \}$, where $e_i \equiv e_j$ denotes that entities $e_i$ and $e_j$ are duplicates or matching, i.e., they describe the same real-world object.
    \item \textit{Dirty ER} or \textit{Deduplication} receives as input a single entity collection, $\mathcal{E}$, with duplicates in itself, and returns as output the set of duplicate entities $\mathcal{D} = \{<e_i, e_j> : e_i \equiv e_j \wedge i \neq j \}$, where $i \neq j$ indicates that $e_i$ and $e_j$ have a different ID.
\end{enumerate}

There is a plethora of research works on Clean-Clean and Dirty ER, leading to a wide range of techniques that span from the seminal rule-based systems and the early probabilistic models to recent machine and deep learning approaches as well as to LLM-based solutions \cite{Papadakis2021morgan,DBLP:journals/pvldb/DongR18,DBLP:conf/www/LiLHZSC24}. Despite this progress, most existing methods share a common underlying assumption: ER is performed as a static decision task, where matches are determined based on fixed representations and pre-defined workflows \cite{DBLP:conf/kdd/GetoorM13,DBLP:books/daglib/0030287}. We call this type of approaches \textit{Passive ER}, since they treat ER as a one-shot computation over a fixed input, without iterative reasoning, adaptive evidence acquisition, or interaction with external sources or users.

More specifically, the dominant framework for Passive ER is the \textit{Filtering–Verification} pipeline \cite{DBLP:journals/corr/abs-2202-12521,DBLP:conf/icwe/NikoletosIP24,Mudgal2018sigmod,DBLP:journals/pvldb/PaulsenGD23}, which structures the end-to-end ER process as a sequence of the four steps in Figure \ref{fig:pipeline}.
The first two steps constitute the Filtering phase, while Matching and Clustering form the Verification phase. Note that we have excluded the Schema Matching step, because we assume highly heterogeneous, continuously ingested input data from a data lake, spanning formats that do not necessarily abide to specific schemata. To address these settings, we assume a \textit{schema-agnostic functionality} of the end-to-end pipeline \cite{DBLP:journals/corr/abs-2304-12329} that disregards schema information in the Blocking and the Meta-blocking, while Matching and Clustering require no aligned schemata.

Recent works on Verification/Matching improve ER effectiveness through deep neural networks in combination with features extracted from pre-trained language models (PLMs) \cite{DBLP:conf/www/ChenSZ20,DBLP:journals/corr/abs-2207-04122,DBLP:conf/edbt/BrunnerS20}. These models produce rich semantic representations of entity records, enabling the matching step to capture subtle lexical and semantic correspondences that surface-level similarity measures miss. However, they are fine-tuned for pairwise classification over serialized record pairs, typically requiring a large set of labelled instances.

To address the lack of labelled instances, more recent matching methods leverage LLMs, reformulating matching as a textual inference problem \cite{DBLP:conf/www/LiLHZSC24,DBLP:conf/coling/WangCLCHSWZ25}: pairs of records are serialized into natural language prompts and fed to an LLM, which outputs a match or non-match decision. This means that LLMs typically serve as zero-shot or few-shot classifiers, which can be fine-tuned with labelled instances for higher performance.



While effective in many settings, Passive ER exhibits fundamental structural constraints: it assumes that all relevant evidence is available a priori, and that matching decisions can be made independently for each pair. Its functionality is typically fixed and non-adaptive, given that most matching methods (e.g., the learning-based ones) do not update their matching decision on new information: they perform one-shot decisions, without supporting multi-step reasoning or evidence acquisition. Only the final Clustering step (if included in the pipeline) partially alleviates this, using global graph knowledge to refine the initial set of matches, but the improvements remain limited \cite{DBLP:journals/vldb/PapadakisETHC23,DBLP:journals/pvldb/HassanzadehCML09}. As a result, Passive ER is not suitable for complex, open-world settings, where resolving ambiguity requires iterative reasoning, adaptive evidence gathering, and cost-aware decision-making.
\section{What is Agentic ER?}
\label{sec:agentic}
In this work, we propose Agentic ER, a new paradigm in which ER is performed by an \textit{autonomous decision-making agent}. Rather than treating ER as a static classification problem, Agentic ER frames it as a \textit{goal-directed, sequential process} in which an agent actively determines how to resolve uncertainty through planning, evidence acquisition, and adaptive decision-making. 

More formally, we define Agentic ER as follows:
\textit{\textbf{\emph{Agentic ER}} is the task of resolving input entity collections by means of autonomous agents that iteratively plan actions, execute them, observe their outcomes, and update their beliefs until stable decisions have been reached.}

\begin{figure}[t]
\centerline{\includegraphics[width=0.4\textwidth]{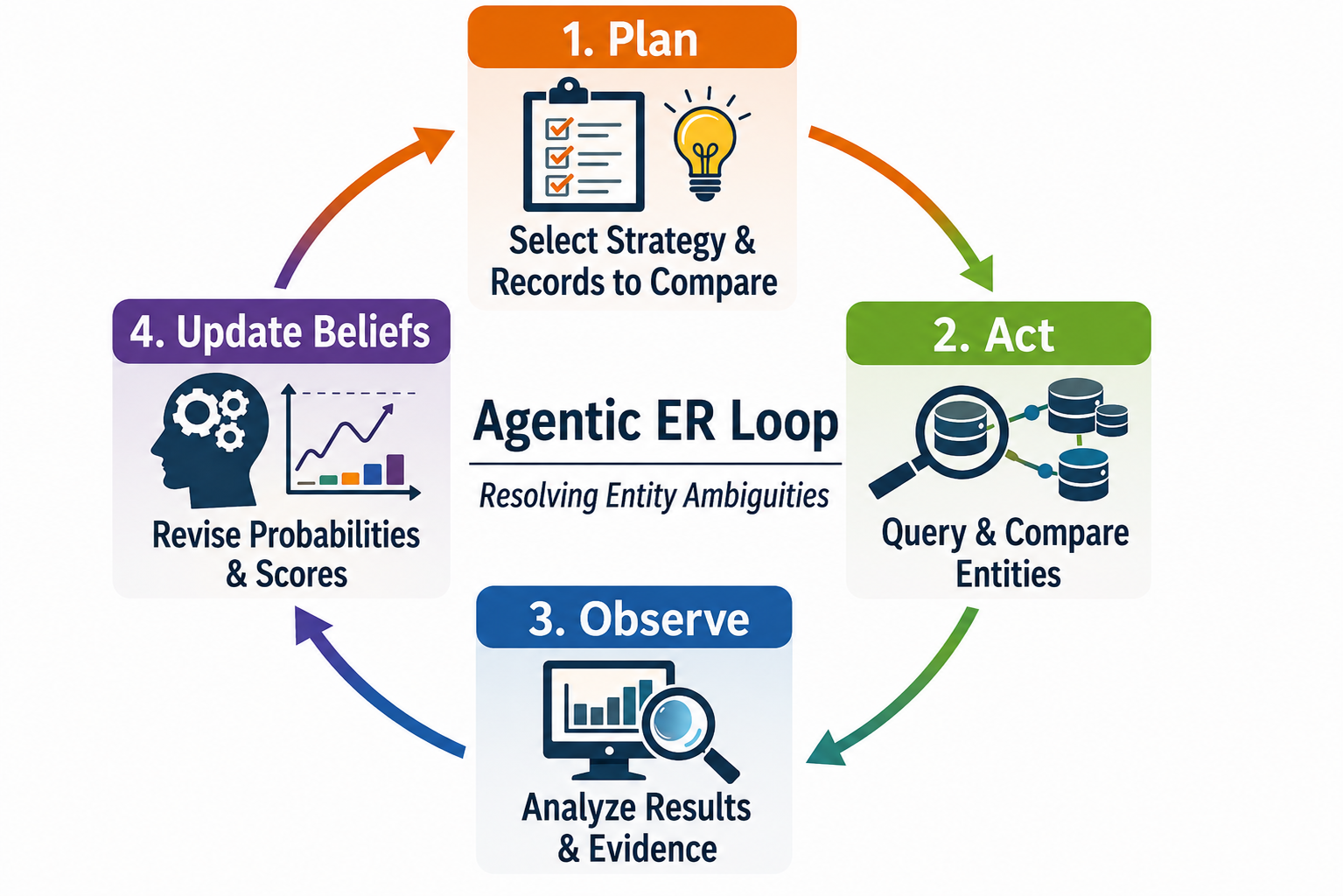}}
\vspace{-8pt}
\caption{The steps lying at the core of Agentic ER.}
\vspace{-14pt}
\label{fig:AERsteps}
\end{figure}

Each agent must determine what to do next in order to reduce uncertainty and achieve the resolution objective. To this end, it operates through a recurring loop with the four steps in Figure \ref{fig:AERsteps}:

\vspace{2pt}
\noindent
\textit{1) Plans.} The agent selects a strategy for resolving an entity ambiguity. This may involve deciding which candidate records to compare first, whether to inspect neighboring records, whether additional evidence is needed, or whether uncertainty is already low enough to stop. Planning can be \textit{explicit}, through a policy or search procedure, or \textit{implicit}, through model-guided action selection.

\vspace{2pt}
\noindent
\textit{2) Acts.} The agent executes actions that modify its information state. Such actions may include comparing a candidate pair, retrieving an external document or profile, querying a knowledge graph, inspecting related records, asking a human for clarification, or merging a set of records into a provisional cluster.

\vspace{2pt}
\noindent
\textit{3) Observes.} Each action yields observations that may confirm, contradict, or refine the current hypothesis. Observations may also include confidence estimations, user responses, or signals from auxiliary ER models.

\vspace{2pt}
\noindent
\textit{4) Updates Beliefs.} Based on the newly acquired observations, the agent revises its internal belief about the matching status of part of the candidate pairs. This belief may be represented as a probability distribution over match hypotheses, a confidence score over clusters, or a structured state that captures partial evidence and unresolved ambiguities.

Overall, Passive ER asks: ``\textit{given one or two entity collections, do their candidate pairs match?}'', whereas Agentic ER asks: ``\textit{given an ambiguous ER task, what sequence of actions should be taken to reach a high-quality decision under resource constraints?}''.

\subsection{Core Characteristics}
\label{sec:characteristics}

The agentic view incorporates the following characteristics that distinguish Agentic ER from Passive ER.

\vspace{2pt}
\noindent
$\bullet$ \textit{Sequential Decision-Making.}
Agentic ER treats ER as a multi-step process rather than a single matching decision. Most importantly, all matching decisions are interdependent for Agentic ER: the outcome of one action depends on the outcomes of previous actions, with future actions being adapted accordingly. This allows Agentic ER to unfold dynamically, especially in difficult or ambiguous cases. 
Note that the iterative approach is automatically determined on-the-fly, during the resolution process.

\vspace{2pt}
\noindent
$\bullet$ \textit{Uncertainty-Aware Reasoning.}
Agentic ER explicitly acknowledges that ER is often uncertain and incomplete. Rather than forcing early binary decisions, the agent maintains and updates beliefs as evidence accumulates. This enables confidence-aware stopping decisions, selective escalation, and more principled trade-offs between accuracy and effort, all carried out automatically.

\vspace{2pt}
\noindent
$\bullet$ \textit{Adaptive Workflows.}
Agentic ER supports adaptive workflows, where the sequence of operations is not predetermined, but is chosen based on the evolving state of the problem. For some cases, a single comparison may suffice. For others, the agent may need to consult multiple sources, inspect related entities, or request external feedback before deciding. This is crucial in data lakes, where new records in heterogeneous formats may arrive continuously, requiring the agent to adapt its resolution workflow on-the-fly.

\vspace{2pt}
\noindent
$\bullet$ \textit{External Knowledge Access.}
Agentic ER naturally lends itself to \textit{open-world settings}, in which not all relevant information is assumed to be present in the input records. An agent may retrieve evidence from web sources, knowledge graphs, enterprise databases, document repositories, or other auxiliary resources. This capability is essential in real settings where entity ambiguity cannot be resolved from local attributes alone.

\vspace{2pt}
\noindent
$\bullet$ \textit{Interaction with Humans or Systems.}
Agentic ER allows the resolution process to be \textit{interactive}. When uncertainty remains high, the agent may ask a human annotator, consult another system, or trigger a domain-specific validation mechanism. Importantly, interaction is not treated as an offline training component only, but as a first-class action that can occur during inference.

\vspace{2pt}
\noindent
$\bullet$ \textit{Cost-Aware Optimization.}
Because actions such as retrieval, querying, or human interaction convey computational, latency, or monetary costs, Agentic ER is framed as a \textit{cost-sensitive process}. Its goal is not only to maximize resolution quality, but to do so efficiently in terms of run-time, memory resources and monetary cost. 

Overall, moving from Passive to Agentic ER transforms the deduplication of entities from a classification task to a \textit{goal-directed process of evidence acquisition and reasoning under uncertainty}.


\subsection{Tool-Selection Policy}
\label{sec:optimizationObjective}

A key component of Agentic ER is the ability to select dynamically among multiple resolution tools and strategies at each step of the pipeline in Figure \ref{fig:pipeline}. We formalize this as a \textbf{tool-selection policy}.

Let $s_t \in \mathcal{S}$ denote the agent's state at time $t$, summarizing its current beliefs, accumulated evidence, and action history. At each step, the agent selects an action $a_t \in \mathcal{A}$ from a rich action space that spans the full ER pipeline. Concretely, the available tools are:
\begin{itemize}[leftmargin=*]
    \item $\mathcal{T}_b$: (meta-)blocking strategies (e.g., syntactic, semantic or hybrid),
    \item $\mathcal{T}_m$: matching models (based on rules, ML, DL, or LLMs),
    \item $\mathcal{T}_r$: retrieval tools (e.g., web search, knowledge graphs),
    \item $\mathcal{T}_h$: human interaction (e.g., annotator queries, expert validation),
    \item $\mathcal{T}_e$: ensemble and voting mechanisms for conflicting signals.
\end{itemize}

\begin{figure}[t]
\centerline{\includegraphics[width=0.47\textwidth]{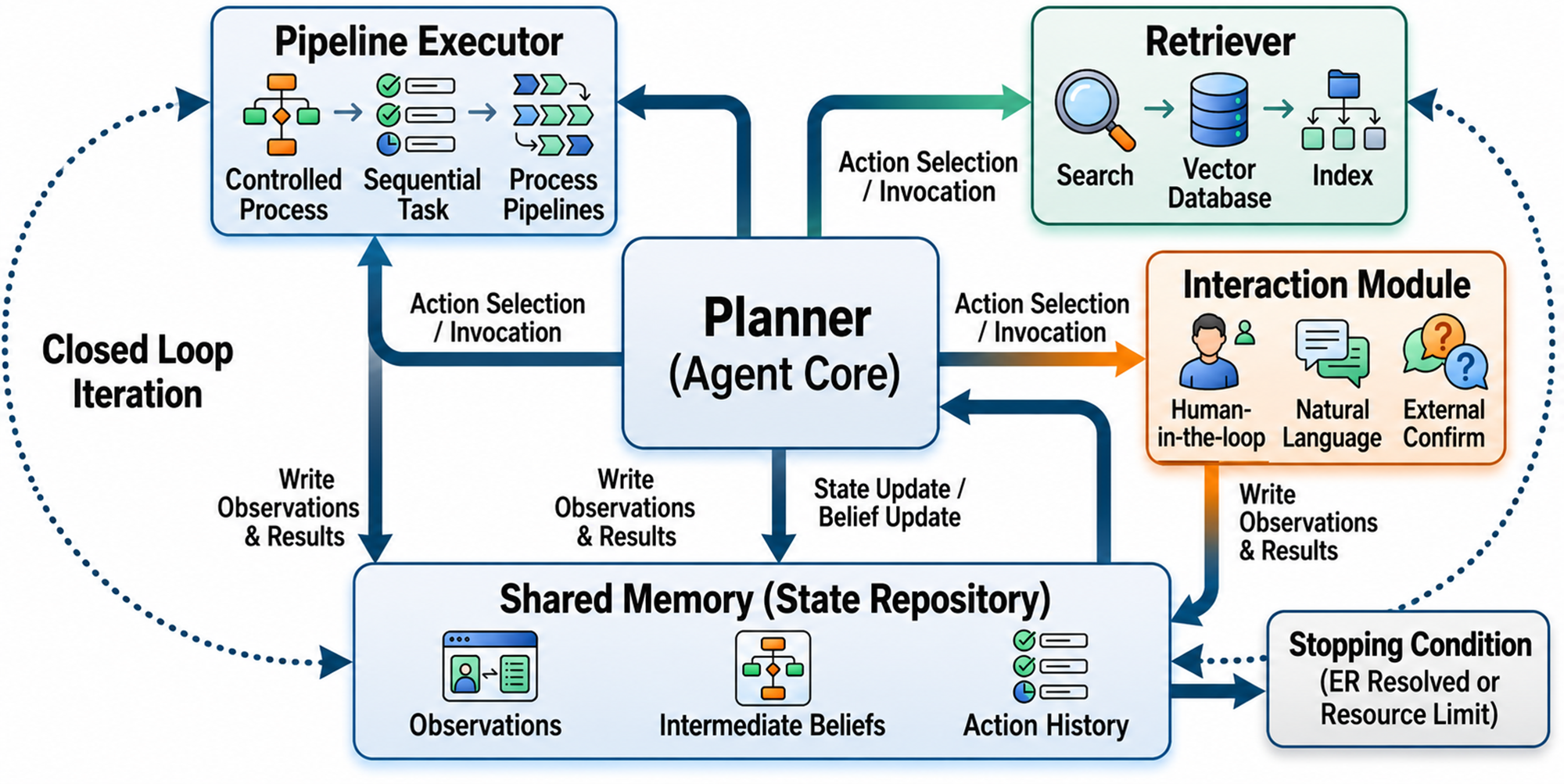}}
\vspace{-8pt}
\caption{Reference for Agentic ER.}
\vspace{-12pt}
\label{fig:architecture}
\end{figure}

Each action is a pair $a_t = (T_i, \theta)$, where $T_i \in \mathcal{T}_b \cup \mathcal{T}_m \cup \mathcal{T}_r \cup \mathcal{T}_h \cup \mathcal{T}_e$ is the selected tool and $\theta$ are its configuration parameters (e.g., matching threshold). The agent's behavior is governed by a \textbf{policy} $\pi(a_t \mid s_t)$ that maps the current state to a distribution over actions.
The policy optimizes the expected trade-off between resolution quality and resource consumption:
$\mathbb{E}\left[ \text{Accuracy} - \lambda_1 \cdot \text{Cost} - \lambda_2 \cdot \text{Latency} \right]$, 
where $\lambda_1, \lambda_2 \geq 0$ control the relative importance of cost and latency. Alternatively, this can be framed as constrained optimization (maximizing accuracy subject to a fixed resource budget), or as cost minimization subject to a target accuracy.

This \textbf{optimization formulation} enables several key capabilities: per-pair strategy selection (e.g., applying a cheap rule-based matcher first and escalating to an LLM only for ambiguous pairs); adaptive blocking before matching; dynamic switching between models based on intermediate confidence; and delegation to ensemble mechanisms for hard cases. This way, Agentic ER becomes a \textbf{meta-decision problem} over ER strategies.
\section{System Architecture for Agentic ER}

To operationalize the Agentic ER paradigm, we propose a high-level system architecture that implements the single-agent model of Section~\ref{sec:agentic} as a coordinated set of specialized modules. Conceptually, these modules constitute the internal components of one Agentic ER agent; in practice, each module can be realized as a sub-agent orchestrated by the Planner, naturally extending to a multi-agent setting as system complexity grows (we identify this direction as an open research challenge in Section~\ref{sec:challenges}). This architecture translates the conceptual and formal models into a system design aligned with the needs of modern data management systems, with each module designed to operationalize a core characteristic from Section~\ref{sec:characteristics}, collectively enabling planning, adaptive evidence acquisition, interaction, and uncertainty-aware reasoning. The proposed architecture comprises five interacting modules, which can be realized using two \textit{complementary} agent orchestration frameworks: \textbf{LangGraph}~\cite{langgraph}, which provides fine-grained programmatic control over model orchestration and tool use, and \textbf{n8n}~\cite{n8n}, which offers low-code integration with external services, messaging systems, and data sources via a visual workflow interface.

At the core of our architecture lies the \textit{Planner}, which is responsible for selecting actions based on the current state or belief. The Planner implements the context-sensitive, decision-making logic of the agent and determines what to do next at each step of the ER process. This logic may be realized through learned policies, such as reinforcement learning, heuristic strategies, or search-based planning methods. For instance, for matching it may apply a cheap similarity function first (e.g., attribute overlap) and escalate to an expensive LLM-based matcher only for pairs that remain ambiguous; for clustering, it decides when to trigger clustering and which algorithm to apply, and may re-run it after new matches are detected. In practice, the Planner can be implemented as an LLM with access to the Memory Module and a registry of available agents; at each step, it determines which agent to invoke and with what objective, receives the output as an observation, and re-enters the Plan--Act--Observe--Update loop.


To support open-world operation, the \textit{Retriever} module accesses external data sources to acquire additional evidence. The Retriever may interface with web search APIs, knowledge graphs, enterprise databases, or document repositories, enabling the agent to resolve ambiguities that cannot be addressed using local data alone. It can be invoked at any pipeline step of Figure \ref{fig:pipeline}. For example, it can augment local record attributes with external information (e.g., alternate names from a knowledge graph) \emph{before} blocking, improving the likelihood that matching records end up in the same block. 
In practice, the Retriever can be implemented as a collection of API endpoints, each tailored to a specific source: for instance, a web search endpoint accepting a natural language query, or a database endpoint issuing SQL queries or API calls against a structured store. This design maps naturally onto n8n, which natively supports connectors for a wide range of external services.


The \textit{Pipeline Executor} is responsible for carrying out any step of the ER pipeline in Figure \ref{fig:pipeline} as directed by the Planner. Rather than executing the full pipeline uniformly over all input entity profiles, the Pipeline Executor is invoked selectively and contextually. Its role varies across the ER pipeline steps in Figure \ref{fig:pipeline}: for blocking, it applies the Planner-selected blocking strategy to the input entity collection(s) and returns the candidate set $C$. For meta-blocking, it applies the selected meta-blocking method (if any) to prune the redundant and superfluous pairs from $C$, returning the refined candidate set $C'$. For matching, it executes the selected matching method over the designated subset of candidate pairs, producing the similarity graph $G$. For clustering, it applies the selected clustering algorithm over the similarity graph $G$, producing the final set of entity clusters $M$. In practice, the Pipeline Executor can be realized either as an LLM-based agent with pipeline tools defined as MCP servers~\cite{fastmcp}, or as a standalone API~\cite{fastapi}. The former allows the model to detect and recover from execution errors (e.g., malformed tool calls) dynamically; the latter offers lower latency and more predictable behavior for well-defined pipeline steps.

Another critical component is the \textit{Memory Module}, which maintains the agent's evolving state. This includes observed evidence, intermediate decisions, belief representations, and the history of actions taken. Note that this module allows the agent to accumulate knowledge over time, supporting sequential reasoning and enabling informed decision-making based on past observations. Note also that it plays a particularly central role in clustering (see Figure \ref{fig:pipeline}), as this is a global operation that depends on the full history of match decisions accumulated across all prior steps. In this context, the agent's memory effectively serves as the evolving similarity graph that clustering operates on. In practice, the Memory Module can be realized using a structured memory library that supports persistent storage of decisions and outcomes alongside semantic search over past events, enabling the Planner to efficiently retrieve relevant prior observations when reasoning about new candidate pairs.

Finally, the \textit{Interaction Module} manages the communication with external agents such as human annotators, domain experts, or auxiliary systems. This module supports the formulation of queries, the handling of responses, and the integration of feedback into the agent's state. Interaction is treated as a first-class operation that can be invoked at any pipeline step: for example, the Planner may consult a domain expert when uncertain about which blocking keys are appropriate, or query a human annotator for high-stakes or persistently ambiguous candidate pairs during matching. In practice, interaction can be triggered directly by the Planner via standard messaging or notification interfaces, and is naturally supported by both LangGraph and n8n.


The interaction among these components realizes the workflow in Figure \ref{fig:AERsteps}. At each step, the Planner selects an action based on the current state stored in memory. The chosen action is then executed by the appropriate module, producing new observations that are incorporated into memory. Based on this updated state, the Planner selects the next action, and the process continues until a stopping condition is met and a final resolution decision is produced.

\section{Benchmarking Agentic ER}

A key barrier to progress in Agentic ER is the lack of appropriate benchmarks and evaluation methodologies. Existing ER benchmarks are crafted for Passive ER solutions and therefore fail to capture the essential properties of agentic behavior. Specifically, they assume that: (i) static evaluation, where all relevant information is provided upfront, (ii) closed-world settings where external evidence is neither required nor modeled, and (iii) no multi-step reasoning is required, incentivizing improvements in similarity modeling rather than advances in planning and decision-making. 

To support the development of Agentic ER solutions, a new generation of benchmarks is needed to explicitly capture the complexity of real-world ER tasks, especially in data lakes.
Such benchmarks should involve:
\begin{enumerate}[leftmargin=*]
    \item \textit{Multi-hop matching.} Resolving a candidate pair requires reasoning across multiple pieces of evidence, potentially involving intermediate outcomes, such as blocking decisions, ambiguous candidate pairs, and conflicting clustering signals. Such tasks naturally fit to sequential reasoning, not to one-shot matching.
    \item \textit{Open-world settings.} Critical information is absent from the input entity profiles, requiring agents to retrieve evidence from external sources at any pipeline step. This setting directly evaluates the ability to perform adaptive evidence acquisition.
    \item \textit{Dynamic and streaming settings.} Entities arrive over time in heterogeneous formats, from structured CSV files to semi-structured JSON and RDF dumps, and agents must incrementally update blocking structures, refine candidate sets, and revise cluster assignments without reprocessing the entire dataset.
\end{enumerate}

Note that using the new benchmark datasets with the above characteristics requires new measures for ER solutions that provide a \textit{multi-dimensional evaluation framework}. The standard measures such as precision, recall, and F1-score remain important for assessing effectiveness, but additional metrics are needed to capture the efficiency and adaptivity of the ER process. These include the total cost incurred (e.g., computation, retrieval, and interaction), the number of actions taken, the interaction overhead, the evidence efficiency (i.e., how much information is required to reach a decision), and the overall latency. Together, these metrics capture the inherent trade-offs in Agentic ER between effectiveness and resource-efficiency as well as adaptivity.
\section{Core Research Challenges}\label{sec:challenges}

Agentic ER introduces a set of open challenges that extend beyond the scope of Passive ER. The following challenges arise from the need to design agents that can plan, act across all steps of the ER pipeline, acquire evidence, interact, and reason under uncertainty, while respecting practical constraints. 

\noindent
\textbf{Planning and Decision-Making.} Unlike Passive ER, where the workflow is fixed, Agentic ER requires the agent to dynamically construct a resolution strategy, deciding which pipeline step to execute next, over which candidate pairs, and with which method. No existing ER system treats the pipeline itself as a space of actions to be planned over. The closest work in the literature is Resolvi \cite{DBLP:journals/corr/abs-2503-08087}, which proposes a reference architecture for ER systems based on a state-based pipeline model, where practitioners determine which components are applicable and in what configuration, rather than following a rigid sequence of execution. However, Resolvi treats pipeline composition as a design-time decision made by human experts, not as a runtime action selected dynamically by an agent. 

Our approach raises the following challenges: (i) \textit{Action selection policies.} How should the agent choose among competing actions such as escalating to an LLM-based matcher, retrieving evidence, querying a human, or triggering clustering? This is quite challenging because the action space spans heterogeneous operation types with vastly different costs and information gains that are difficult to estimate a priori. (ii) \textit{Sequential dependencies.} The actions of the agents are interdependent, given that early decisions affect the later ones. 
(iii) \textit{Search vs. Learning.} Should planning be performed via explicit search or learned policies (e.g., reinforcement learning)? The former requires large amounts of labeled ER trajectories that do not currently exist, while the latter struggles to scale to the large candidate spaces of real-world ER. (iv) \textit{Scalability.} How can planning remain efficient in large datasets with many candidate entities and actions? To the best of our knowledge, no work in the literature addresses these new ER challenges.

\noindent
\textbf{Evidence Acquisition.} A defining feature of Agentic ER is the ability to operate in \textit{open-world settings}, where not all relevant information is available upfront. Thus, the Retriever can be invoked at any pipeline step to augment the available information. 
This introduces the following challenges: (i) \textit{When to retrieve.} How should the agent decide whether current evidence is sufficient? The challenge is to estimate the expected value of additional evidence before it is acquired, under uncertainty about what that evidence will contain. (ii) \textit{What to retrieve.} Which external sources are most likely to reduce uncertainty at a given pipeline step? This is challenging because the relevance of a source depends on the specific ambiguity at hand, which varies per candidate pair and pipeline step. (iii) \textit{How much to retrieve.} How can the agent avoid excessive or redundant retrieval? The challenge is to balance over-retrieval, which introduces noise and cost, with under-retrieval, which leaves ambiguities unresolved. (iv) \textit{Evidence integration.} How should heterogeneous and potentially conflicting evidence be aggregated? A form of evidence acquisition has been applied to ER in \cite{DBLP:journals/corr/abs-2510-14271}, which leverages RAG-based ER for reducing noise in knowledge graphs; however, retrieval there is a fixed pre-processing step rather than an adaptive, agent-controlled action, leaving the full evidence acquisition problem open. Overall, none of these challenges has been addressed in the ER literature.

\noindent
\textbf{Cost-Aware Optimization.}
In Agentic ER, actions such as running an LLM matcher, retrieving external evidence, or querying a human incur \textit{non-negligible costs}. This necessitates a shift from effectiveness-oriented optimization to \textit{cost-aware decision-making}. The challenge is to jointly optimize quality, latency, and monetary cost over a dynamic, variable-length sequence of actions. We break this core challenge into the following fine-grained ones: (i) \textit{Trade-off Modeling.} How can agents balance effectiveness, latency, and monetary or computational cost across pipeline steps? Static trade-off models are inapplicable, because the cost and quality contribution of each action depend on the current resolution state. (ii) \textit{Budget-Constrained Resolution.} How should the agent operate under fixed resource budgets? This is challenging because budgets may be exhausted mid-pipeline, requiring the agent to commit to a decision with incomplete evidence. (iii) \textit{Adaptive Stopping.} When should the agent stop gathering evidence and commit to a decision? The challenge is to avoid stopping too early, yielding incorrect matches, and too late, wasting resources. (iv) \textit{Cost Heterogeneity.} Different actions have different costs (e.g., querying a human vs. running a trained AI model) or are uncertain (e.g., human response time is variable). The most relevant works in the literature pertain to cost-aware solutions for crowd-sourced ER, e.g., \cite{DBLP:journals/vldb/ChaiLLDF18}. However, they are crafted for a static task allocation problem, rather than a dynamic task that interleaves human queries with automated pipeline steps.

\noindent
\textbf{Human-in-the-Loop Interaction.}
The Interaction Module enables human expertise to be injected at any pipeline step, from selecting blocking keys to validating ambiguous candidate pairs or suspicious clusters. As explained above, the Retriever specifies \textit{when} to ask for help, while existing solutions for crowd-sourced ER specify \textit{what} to ask and \textit{how} to ask it, as well as how to minimize interaction cost \cite{Papadakis2021morgan}. The main challenge for our Interaction Module is \textit{learning from feedback}: how can the agent incorporate feedback to improve future decisions? This is non-trivial because inference-time feedback arrives incrementally and out of order, making it difficult to update the agent's policy without destabilizing prior decisions. 
Similar ideas
have been explored in \cite{DBLP:conf/icdm/Christen15,DBLP:conf/cikm/GurajadaPQS19,DBLP:conf/sigmod/ChandraHKR94}, but they treat interaction as an offline or training-time activity.

\noindent
\textbf{Uncertainty and Explainability.}
Uncertainty is inherent in ER, while explainability constitutes a core requirement for ER systems. Works on uncertainty-aware ER are summarized in \cite{DBLP:journals/pvldb/Gal14}, while solutions to explainable ER are discussed in \cite{DBLP:journals/pvldb/QianPS19,DBLP:journals/tkde/Barlaug23,DBLP:conf/icde/EbaidTAEO19}. All these works, however, focus on Passive ER, where a single model (typically a binary classifier) produces individual matching decisions. As a result, these works cannot be extended to the dynamic, multi-step process of Agentic ER, which applies to open-world and ambiguous settings. In these settings, the Memory Module accumulates potentially noisy or conflicting evidence across multiple runs. Agentic ER must therefore explicitly address both uncertainty modeling and explainability in an entirely new context.
More specifically, the key challenges are: (i) \textit{Confidence Estimation.} How can the agent quantify uncertainty in match decisions and intermediate states? This is challenging because uncertainty must be estimated not just over individual matching pairs, but over partial, evolving resolutions that depend on a history of actions and observations. (ii) \textit{Uncertainty Propagation.} How does uncertainty evolve as blocking, matching, and clustering decisions accumulate? This is challenging because errors introduced in early pipeline steps compound in later ones in ways that are difficult to track analytically. Note that all existing works focus on uncertainty in the matching step, independently of the preceding blocking phase \cite{DBLP:journals/pvldb/Gal14}. (iii) \textit{Decision Justification.} How can the agent provide interpretable explanations for its full action history? This is challenging because in Agentic ER, the final decision results from a long sequence of heterogeneous actions (i.e., pipeline steps, retrievals, human queries); summarizing them into a human-readable justification requires novel explanation generation techniques. (iv) \textit{Trust and Accountability.} How can users trust a resolution process that unfolds across multiple steps and external sources? This is challenging because users must be able to audit not just the final decision but the entire reasoning trace, including retrieved evidence, posed queries, and points of uncertainty. 

Note that these challenges are deeply interconnected: planning depends on cost models and uncertainty estimates, evidence acquisition influences both accuracy and cost, and interaction strategies rely on value-of-information considerations.

\vspace{-12pt}
\section{Conclusions}
\vspace{-5pt}

In this paper, we argued for a paradigm shift in ER: from passive, one-shot matching to agentic, sequential decision-making. We introduced Agentic ER as a unifying framework in which autonomous agents plan actions, acquire evidence across all steps of the ER pipeline, interact with external sources, and optimize trade-offs between accuracy and cost. We provided a formalization of Agentic ER, we highlighted its distinctive characteristics, we proposed a modular reference architecture, and we identified key research challenges and benchmarking directions. By tackling these challenges, we can move toward ER systems that are adaptive, interactive, and resource-aware, thus being capable of operating effectively in complex, real-world environments such as data lakes with streaming, heterogeneous content spanning CSV files, JSON files, RDF dumps, knowledge graphs, and free text. We believe that Agentic ER opens a rich design space at the intersection of data management and intelligent agents, representing a step toward intelligent data systems that can reason, interact, and adapt. 

\bibliographystyle{ACM-Reference-Format}
\bibliography{refs}

\end{document}